# Numerical Simulation of the Impact of Different Cushion Gases on Underground Hydrogen Storage in Aquifers Based on an Experimentally-Benchmarked Equation-of-State

Qingqi Zhao[1], Yuhang Wang[2], Cheng Chen[1]*

[1] Department of Civil, Environmental and Ocean Engineering, Stevens Institute of Technology, Hoboken, NJ, USA

[2] School of Environmental Studies, China University of Geosciences (Wuhan), Wuhan, China

* Corresponding author: Cheng Chen (cchen6@stevens.edu)

**Abstract**

Underground hydrogen storage (UHS) in geological formations is a promising technology for large-scale hydrogen energy storage. In this study we focus on UHS in saline aquifers. Although lessons were learned from similar studies, including geological carbon sequestration and underground gas storage, the unique thermodynamic and physical properties of hydrogen distinguish UHS from the other subsurface storage projects. We developed a two-phase, three-component reservoir simulator, which incorporated essential physics based on the fully coupled multi-physics framework of the Delft Advanced Reservoir Simulation (DARSim). Properties of fluid mixtures were computed using the GERG-2008 equation of state (EoS), in which the parameters were determined by fitting laboratory data. The simulation results demonstrated the impact of different cushion gases ($CO_2$, $CH_4$, and $N_2$) on UHS. Most hydrogen stayed in the gaseous phase at the top of the aquifer when $CH_4$ and $N_2$ were used as the cushion gas. Conversely, hydrogen-rich fingers were observed in the aqueous phase when $CO_2$ was used as the cushion gas, because dissolved $CO_2$ increased brine density, leading to density-driven downward convection which was favorable for hydrogen dissolution into the aqueous phase. The highest purity of produced hydrogen was observed when $CO_2$ was used as the cushion gas, whereas using $CH_4$ and $N_2$ as the cushion gas was favorable for the hydrogen production rate and mobility. This work is the first study that utilizes an EoS-based reservoir simulator to investigate hydrogen's flow patterns and interactions with cushion gases in an underground storage system. Specifically, the productivity and retrievability of hydrogen in a two-phase, three-component UHS system were analyzed in detail. The developed reservoir simulation tool and research findings from this study will be valuable to support decision making in practical UHS projects.

## 1. Introduction

Large-scale consumption of fossil energy has resulted in global climate changes, air pollution, and energy crisis. Hydrogen, as a clean energy source, has the potential to replace fossil fuels because of its zero emission and relatively high specific energy capacity (120-142 kJ/g) [1,2]. Furthermore, hydrogen can be served as an efficient energy carrier, which has a wide range of applications in various industries such as hydrogen-powered vehicles and steel production [3–5].

According to the International Energy Agency report, the global hydrogen demand is expected to reach 528 million metric tons by 2050 [6]. Successful application of hydrogen energy at large scale depends primarily on efficient storage and transportation technologies at large scale. Underground hydrogen storage (UHS) is a feasible technology to store mega tons of hydrogen (i.e., energy storage at the terawatt-hour scale), allowing for balancing the fluctuating renewable energy production to reliably supply the fluctuating consumption. More precisely, UHS is a promising approach compared with surface-based storage tanks due to its ability to store large-scale hydrogen for extended periods, minimum occupancy of surface space, and natural isolation of hydrogen from oxygen [7]. The knowledge and experience gained from underground gas storage (UGS) can also be transferred to UHS, in some aspects, specially the geomechanics and storage volume estimates [8]. With a history dating back to 1915, UGS had established 680 facilities with a total storage capacity of 413 billion cubic meters as of 2015 [9].

The use of subsurface saline aquifers as the UHS site is feasible due to their widespread distributions around the world. A subsurface saline aquifer contains a permeable porous medium filled with brine, and is covered by an impermeable cap rock [2]. Hydrogen is injected into a subsurface saline aquifer and displaces the in-situ brine in lateral and vertical directions. It should be noted that during the production stage, brine can be produced with hydrogen simultaneously, which is unfavorable for UHS in saline aquifers [9–11]. No practical UHS projects in saline aquifers have been reported so far, which emphasizes the necessity for further research on the feasibility of UHS in subsurface saline aquifers.

In a typical UHS system, hydrogen is withdrawn from the saline aquifer in a cyclic manner, resulting in a loss of reservoir pressure. Therefore, a cushion gas, such as $CO_2$, $CH_4$, and $N_2$, is needed to maintain the reservoir pressure and production flow rate. However, injection of a cushion gas into the formation may lead to lower recovery rate and purity of produced hydrogen. Extraction of brine from the saline aquifer also contributes to the purity reduction of produced hydrogen [2]. The low purity of produced hydrogen may produce negative effects for UHS projects, including toxicity, explosion issues, and increased workload for compression and dehydration. Although these issues may be alleviated over time, they must be considered carefully in the design and planning phase.

The characteristics of multi-component fluid interfaces and fluid mixing interactions affect the overall performance of a UHS system. Therefore, beside the hydrodynamics properties of hydrogen transport in porous aquifers, it is also critical to thoroughly investigate the thermodynamic properties (e.g., viscosity, density, compressibility) when injected hydrogen mixes with the cushion gas in an UHS system. Hydrogen density in saline aquifers is much lower than other cushion gases and brine. The buoyancy force drives hydrogen to the top of the aquifer, covered by a tight, low-permeability cap rock. The density of hydrogen increases with pressure, allowing for more hydrogen storage at deeper depths. This increase, however, is not as significant as for $CO_2$, since $CO_2$ turns into super critical dense fluid below approximately 800 m of depth [8]. In addition, the large difference between hydrogen and brine viscosities impedes hydrogen drainage and can lead to uncontrolled lateral spreading of hydrogen [12], causing a reduction in displacement efficiency. Specifically, compared with cushion gases like $CO_2$, hydrogen viscosity is not sensitive to temperature or pressure under the reservoir conditions [13].

On the other hand, the solubility of hydrogen in brine decreases with temperature and salinity but increases with pressure. Models and correlations have been developed to accurately calculate hydrogen solubility in water and brine [14]. Due to the nonpolar molecular structure of hydrogen, its solubility in pure water is only 0.00103 mol/mol at 323 K and 7.9 MPa, which is similar to $CH_4$ but one order of magnitude lower than $CO_2$. Therefore, The influence of hydrogen solubility on the pore fluid's pH value can be ignored and the dissolution trapping mechanism is not significant for hydrogen [13,15].

Existing thermodynamic models can accurately predict the properties of pure hydrogen. However, the calculation of hydrogen properties in multi-phase and multi-component systems, particularly under high temperature and pressure conditions in a subsurface geological formation, is uncertain. This is partially due to limited experimental data on hydrogen when it is mixed with other fluids, which hinders the validation and tuning of thermodynamic models. Recently a molecular thermodynamics approach has been developed for interfacial properties of hydrogen-brine, which is also validated against the sparsely available experimental data [16]. In addition to these interfacial properties, an accurate equation of state (EoS) is indeed required to describe the properties of hydrogen mixtures. The GERG-2008 EoS [17] has been validated as a robust model for predicting the thermodynamic properties of hydrogen when it is mixed with other gas components over a wide range of temperatures and pressures (i.e., 60 to 700 K and up to 70 Mpa). Beckmüller et al. [18] proposed a new EoS based on the GERG-2008 framework to predict the properties of hydrogen mixed with $CH_4$, $N_2$, CO, and $CO_2$. Hassanpouryouzband et al. [19] predicted the thermodynamic properties of hydrogen mixed with $CH_4$, $N_2$, $CO_2$, and a group of natural gas combinations, and established a comprehensive database covering a wide range of temperatures and pressures. Alanazi et al. [20]evaluated the performances of the cubic, perturbed-chain statistical associating fluid theory (PC-SAFT), and GERG-2008 EoS in calculating the thermophysical properties of gas mixtures containing hydrogen. However, the GERG-2008 EoS needs to be modified to account for the presence of a water-rich aqueous phase [17]. Chabab et al.[21] modified the Soreide and Whitson model with new parameters which were tuned based on experimental data of hydrogen, $CH_4$, $N_2$, and $CO_2$. The proposed model can predict the solubility of gas mixtures up to two gas components in a two-phase system. Alanazi et al. [22] investigated the Schwartzentruber and Renon modified Redlich–Kwong cubic EoS and PC-SAFT in hydrogen solubility in water. This model can predict hydrogen solubility in water within a binary gas mixture system.

Several studies have been conducted on reservoir simulations for UHS. Zamehrian and Sedaee [23] studied UHS in a depleted natural gas reservoir with different cushion gases. They concluded that the $N_2$ had the best performance when used as a cushion gas because it led to the highest amount and purity of produced hydrogen. Eller et al. [24] modeled hydrogen flow in an aquifer coupled with the PC-SAFT EoS; they found an agreement between calculated transport properties and experimental data. Wang et al. [25] investigated hydrogen storage in subsurface porous media with $CO_2$ as the cushion gas; they found that the injected $CO_2$ volume increased by 30% when $CO_2$ dissolution in brine is considered. Cai et al. [26] developed the GPSFLOW simulation tool for modeling hydrogen and gas mixtures in a subsurface storage site, which is applicable for both salt caverns and depleted gas reservoirs. Bo et al. [27] simulated the cyclic hydrogen storage in an Australian gas field, based on experimentally-validated relative and capillary pressure functions. They concluded, among others, that the hysteretic behavior of the hydrogen transport is a key factor in its recovery factor estimate. Pan et al. [28] analyzed the impact of relative permeability hysteresis, rock wettability, and injection and withdrawal schemes on the hydrogen withdrawal efficiency. The results showed that hydrogen-brine relative permeability hysteresis led to higher hydrogen withdrawal purity; however, more water-wet conditions caused a lower hydrogen withdrawal efficiency and a higher water production. Heinemann et al. [29] studied the effects of hydrogen injection and production operations on UHS performance. They found that operation strategies, including well configurations and operational schedules, played a significant role in the UHS project. **Table 1** gives a comprehensive literature review that summarizes the recent studies on UHS reservoir simulations.

**Table 1.** Literature review of UHS reservoir simulations.

| Studies | Storage formation | Cushion gases | Consider relative permeability hysteresis? | Compare different injection and withdrawal operations? | Consider hydrogen dissolution in water? |
|---|---|---|---|---|---|
| Pfeiffer [10] | Sandstone reservoir | $N_2$ | No | No | No |
| Feldmann [30] | Depleted gas reservoir | $N_2$, $CH_4$ | No | No | No |
| Sainz-Garcia [11] | Aquifer | None | No | Yes | No |
| Lubon and Tarkowski [31] | Aquifer | None | No | No | No |
| Mahdi et al. [32] | Sandstone reservoir | None | Yes | Yes | No |
| Cai et al. [26] | Aquifer, depleted gas reservoir | $N_2$, $CH_4$, $CO_2$ | No | No | Yes |
| Kanaani et al. [33] | Depleted oil reservoir | $N_2$, $CH_4$, $CO_2$ | No | Yes | Yes |
| Okoroafor et al. [34] | Depleted gas reservoir | $CH_4$ | No | Yes | Yes |
| Zamehrian and Sedaee [23] | Depleted gas condensate reservoir | Natural gas, $N_2$, $CH_4$, $CO_2$ | No | Yes | No |
| Eller et al. [24] | Aquifer | None | No | No | Yes |
| Wang et al. [25] | Aquifer | $CO_2$ | No | Yes | No |
| Bo et al. [27] | Depleted gas reservoir, underlain by an aquifer | None | Yes | Yes | No |
| Pan et al. [28] | Aquifer | None | Yes | Yes | No |
| Heinemann et al. [29] | Aquifer | Cushion gas type not mentioned | No | Yes | No |
| Lysyy et al. [35] | Aquifer | $N_2$ | Yes | No | No |
| Chai et al. [36] | Aquifer | $N_2$, $CH_4$, $CO_2$ | Yes | No | Yes |

Despite the numerous reservoir simulation studies mentioned above, simulating multi-phase multi-component UHS reservoirs remains a challenge due to the complexities of hysteretic transport characteristics, dissolution of hydrogen in brine, and gas mixtures hydro-thermodynamics. Nonhysteretic relative permeability can overestimate cumulative hydrogen production [27,28,35]. This is because gas trapping reduces the available pore space for the injected hydrogen. Conversely, when considering the hysteresis of $H_2$-brine relative permeability, it could lead to higher purity in the produced hydrogen [28]. The challenges related to dissolution and mixture thermodynamics can be resolved by accurate modeling of the mixtures' thermodynamic properties such as density and viscosity. In this study, we develop such an accurate modeling technique. To this end, a two-phase, three-component reservoir simulator was developed, which incorporates essential physics based on the fully-coupled multi-physics framework of the multiscale DARSim simulator [37]. Fluid properties were calculated using the GERG-2008 EoS with optimized parameters based on experimental data from literature.

The paper is structured as follows. First, we introduce the GERG-2008 EoS for the calculation of thermodynamic properties of the hydrogen-cushion gas mixture. Next, we discuss the governing equations and numerical modeling methods. The numerical model is then validated by comparison with a multi-component flow model in the literature. The reservoir geometry and operational strategy in the case study are then described. Simulated mole fraction distribution in different phases is demonstrated; the production rate, dissolution, and purity of produced hydrogen are quantified and compared between different cushion gas scenarios.

## 2. Methods

### 2.1 GERG-2008 EoS and model calibration

To understand the fluid displacement pattern in UHS, precise predictions of thermodynamic properties using an accurate EoS are critical. In this study, the GERG-2008 EoS, with parameters tuned against experimental data, was used to calculate the fluid properties of gas mixtures and the gas solubility in water in UHS.

GERG-2008 EoS is explicit in the Helmholtz free energy and is a function of temperature $T$, density $\rho$ (representing pressure), and molar composition $x$. A detailed description of the mathematical correlations for different thermodynamic properties of the fluids has been given in the literature [17]. In general, the dimensionless form of the Helmholtz equation is:

$$\frac{a(\rho,T,x)}{RT} = \alpha(\delta, \tau, x) \tag{1}$$

where $R$ is the molar gas constant, $\alpha$ is the reduced Helmholtz free energy, $\delta$ is the reduced mixtures density, and $\tau$ is the inverse reduced temperature. Furthermore, the Helmholtz energy consists of two parts, including an ideal part, $\alpha^0$, which denotes the ideal fluid, and a residual part, $\alpha^r$, which denotes the interactions between real fluids:

$$\alpha(\delta, \tau, x) = \alpha^0(\rho, T, x) + \alpha^r(\rho, T, x) \tag{2}$$

The ideal fluid term of the Helmholtz free energy in the dimensionless form is described as

$$\alpha^0(\rho, T, x) = \sum_{i=1}^{N} x_i [\alpha_{o,i}^o(\rho, T) + \ln x_i] \tag{3}$$

where $\alpha_{o,i}^o$ and $x_i$ is the ideal-gas state Helmholtz free energy and mole fraction of the pure component i in dimensionless form, respectively. $N$ is the total number of components. The term $x_i \ln x_i$ describes the entropy of the mixtures.

The residual term of the Helmholtz free energy is given by

$$\alpha^r(\delta, \tau, x) = \sum_{i=1}^{N} x_i \alpha_{o,i}^r(\delta, \tau) + \Delta \alpha^r(\delta, \tau, x) \tag{4}$$

where $\alpha_{o,i}^r$ is the residual term of the reduced Helmholtz free energy for component i, and $\Delta \alpha^r$ is the departure term of the mixtures. The departure term depends on the reduced temperature $\tau$, the reduced density $\delta$, and the mixture composition. The reduced density, $\delta$, is calculated as

$$\delta = \frac{\rho}{\rho_r(x)} \tag{5}$$

where $\rho_r$ is the reduced mixture density and calculated as

$$\frac{1}{\rho_r(x)} = \sum_{i=1}^{N} x_i^2 \frac{1}{\rho_{c,i}} + \sum_{i=1}^{N-1} \sum_{j=i+1}^{N} 2 x_i x_j \beta_{v,ij} \gamma_{v,ij} \times \frac{x_i + x_j}{\beta_{v,ij}^2 x_i + x_j} \frac{1}{8} \left( \frac{1}{\rho_{c,i}^{\frac{1}{3}}} + \frac{1}{\rho_{c,j}^{\frac{1}{3}}} \right)^3 \tag{6}$$

The reduced temperature, $\tau$, is calculated as

$$\tau = \frac{T_r(x)}{T} \tag{7}$$

where $T_r$ is the reduced mixture temperature and is defined as

$$T_r(x) = \sum_{i=1}^{N} x_i^2 T_{c,i} + \sum_{i=1}^{N-1} \sum_{j=i+1}^{N} 2 x_i x_j \beta_{T,ij} \gamma_{T,ij} \times \frac{x_i + x_j}{\beta_{T,ij}^2 x_i + x_j} \sqrt{T_{c,i} T_{c,j}} \tag{8}$$

The binary parameters $\beta_{v,ij}, \gamma_{v,ij}, \beta_{T,ij}$, and $\gamma_{T,ij}$ are determined by fitting experimental data. The relationship between binary parameters and the order of the components is described as follows:

$$\beta_{v,ij} = \frac{1}{\beta_{v,ji}} \tag{9}$$

$$\beta_{T,ij} = \frac{1}{\beta_{T,ji}} \quad (10)$$

$$\gamma_{v,ij} = \gamma_{v,ji} \quad (11)$$

$$\gamma_{T,ij} = \gamma_{T,ji} \quad (12)$$

The GERG-2008 EoS does not include the calculation of dynamic viscosity. The mixture's viscosity is calculated by a residual entropy scaling approach based on the GERG-2008 EoS framework [38]. The viscosity $\eta$ is calculated as:

$$\eta = \eta^* \times \eta^{ref} \quad (13)$$

where $\eta^{ref}$ is the reference viscosity, and $\eta^*$ is the dimensionless viscosity. The reference viscosity $\eta^{ref}$ can be calculated using the Chapman−Enskog viscosity, $\eta^{CE}$:

$$\eta^{CE} = \frac{5}{16} \frac{\sqrt{Mk_BT/(N_A\pi)}}{\sigma^2 \Omega^{(2,2)^*}} \quad (14)$$

where $M$ is the molecular mass, $k_B$ represents the Boltzmann's constant, $N_A$ is the Avogadro's number, $\sigma$ is the Lennard-Jones segment diameter, and $\Omega^{(2,2)^*}$ is the collision integral. Therefore, the dimensionless viscosity is calculated as

$$\eta^* = \frac{\eta}{\eta^{CE}} \quad (15)$$

The logarithm, $ln\eta^*$, is calculated using a third-order polynomial of the molar residual entropy, $s^{res}$:

$$ln\eta^* = A + \frac{Bs^{res}}{R} + C\left(\frac{s^{res}}{R}\right)^2 + D\left(\frac{s^{res}}{R}\right)^3 \quad (16)$$

where A, B, C, and D are the optimized coefficients determined by experimental data.

The GERG-2008 EoS has been validated in previous studies for gas mixtures of $CH_4$, $CO_2$, and natural gas [39–44]. **Figure 1** illustrates the validation of the GERG-2008 EoS code developed in this work against experimental data [45,46]. Calculation of density was verified at two temperatures, 323 K and 348 K, with a hydrogen mole fraction of 26% and a $CH_4$ mole fraction of 74%. The viscosity validation was performed at temperatures of 323 K and 363 K with a hydrogen mole fraction of 30% and a $CH_4$ mole fraction of 70%. The results of the comparison demonstrate the high accuracy of the GERG-2008 EoS code developed in this study in predicting the thermodynamic properties of gas mixtures.

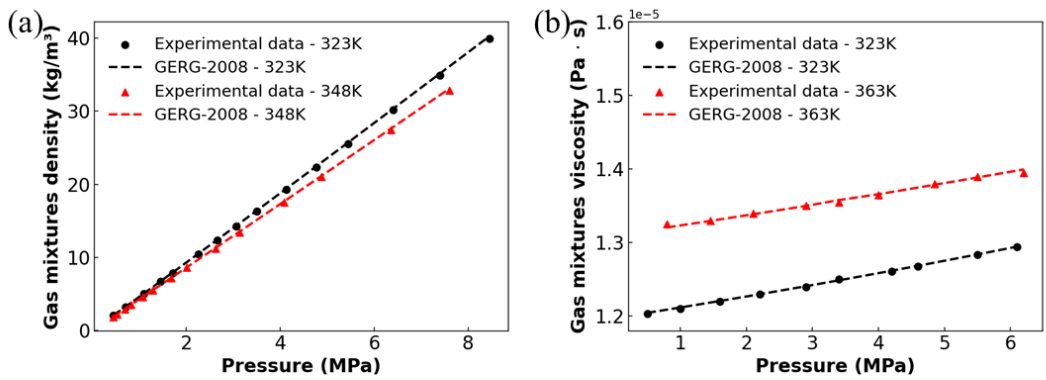

**Figure 1.** GERG-2008 EoS calculation and experimental data [45,46] for a) density of a gas mixture of 26% $H_2$ and 74% $CH_4$, and b) dynamic viscosity of a gas mixture of 30% $H_2$ and 70% $CH_4$ under different temperatures and pressures.

**Figure 2** presents the density and viscosity of gas mixtures of hydrogen and three different cushion gases ($CO_2$, $CH_4$, and $N_2$) calculated using the GERG-2008 EoS. The properties were calculated under the varying temperature from 338 to 343 K with increasing pressure with a fixed mole fraction combination of 50% hydrogen and 50% cushion gas. The density of the mixture is observed to decrease with an increase in temperature, while the viscosity displays an opposite trend, increasing with temperature, with the exception of the hydrogen-$CO_2$ mixtures at high pressures. This deviation may be attributed to the supercritical behavior of $CO_2$ in these conditions.

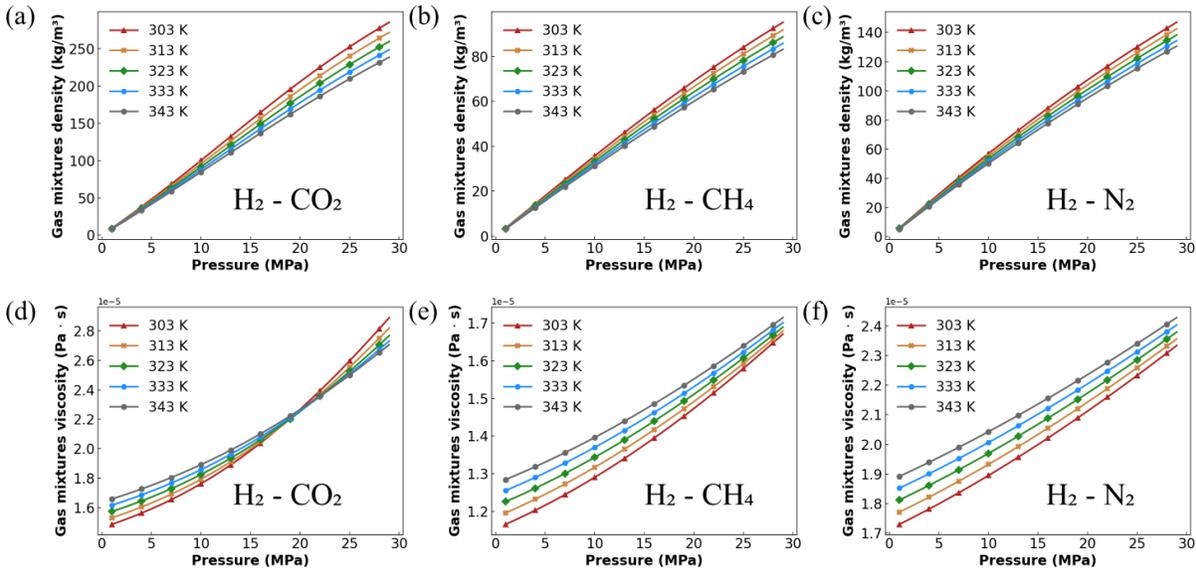

**Figure 2.** Densities of a) $H_2$ - $CO_2$, b) $H_2$ - $CH_4$, and c) $H_2$ - $N_2$ mixtures, and dynamic viscosities of d) $H_2$ - $CO_2$, e) $H_2$ - $CH_4$, and f) $H_2$ - $N_2$ mixtures, under various temperatures and pressures. The mole fraction combination is fixed at 50% hydrogen and 50% cushion gas. These calculations were conducted using the GERG-2008 EoS.

The GERG-2008 EoS has been proven to have a high accuracy [17] in predicting thermodynamic properties of the mixtures of hydrogen and natural gas. However, the model needs to be improved when considering the aqueous phase. Therefore, to accurately predict the dissolution of hydrogen and other gases in the aqueous phase, the mixture interaction parameters, $\beta_{v,ij}, \gamma_{v,ij}, \beta_{T,ij}, \gamma_{T,ij}$ in Equations. (7) – (8), are determined through data calibrating against experimental data [47–50]. The calibrated parameters are illustrated in **Table 2**.

**Table 2.** Calibrated binary parameters for the calculations of reduced density and temperature.

| $i + j$ | $\beta_{v,ij}$ | $\gamma_{v,ij}$ | $\beta_{T,ij}$ | $\gamma_{T,ij}$ |
|---|---|---|---|---|
| $H_2$ + water | 1.0 | 1.0 | 0.88514 | 3.45786 |
| $CH_4$ + water | 1.10278 | 1.0 | 0.87670 | 1.29757 |
| $N_2$ + water | 1.0 | 1.09475 | 1.38500 | 0.96881 |
| $CO_2$ + water | 0.8 | 0.8 | 1.30237 | 0.79545 |

To validate the accuracy of the GERG-2008 EoS with calibrated parameters for the prediction of gas solubility in water, a validation experimental dataset [12,51–55] was gathered. **Figure 3** illustrates the GERG-2008 EoS-calculated solubility of hydrogen, $CH_4$, $N_2$, and $CO_2$ in water with calibrated parameters, in comparison with experimental data from the validation dataset. Overall, it was observed that the solubility of $CO_2$ is one order of magnitude higher than other gases, whereas $N_2$ has the lowest solubility.

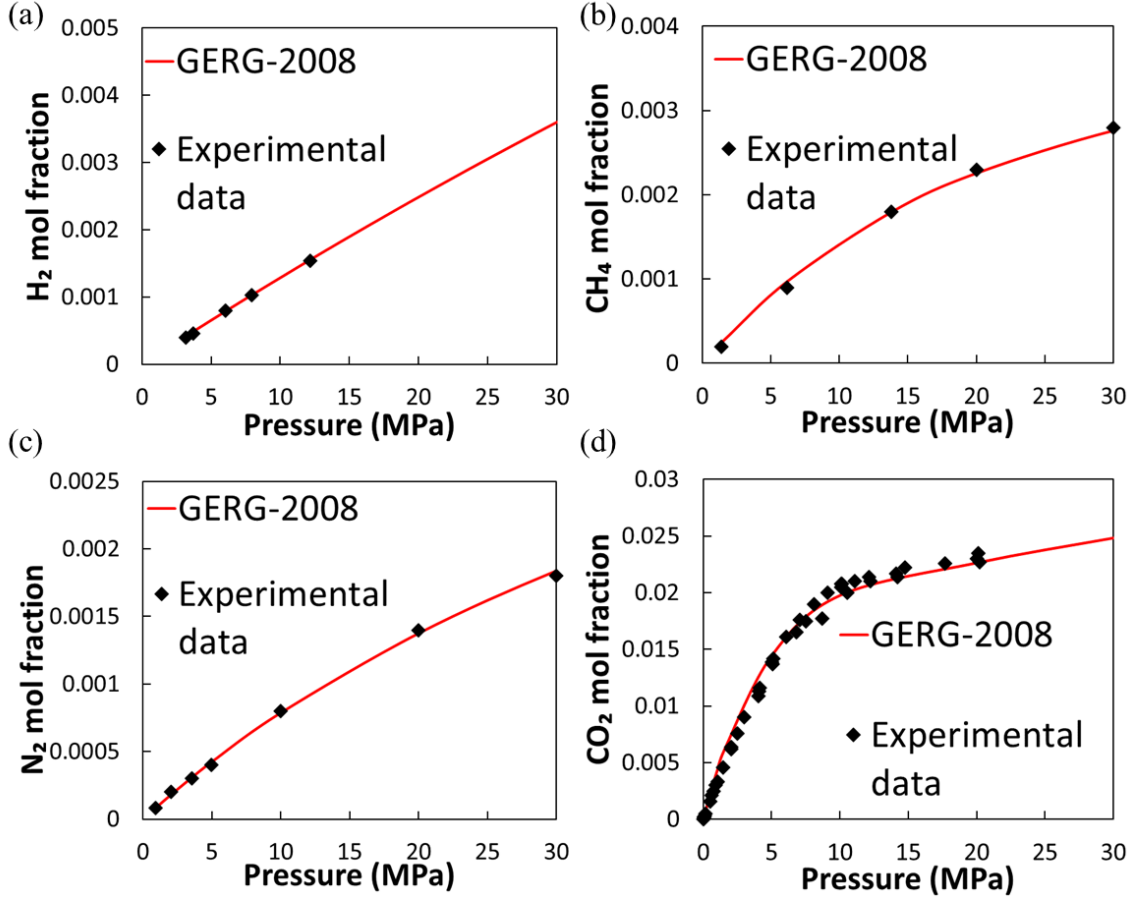

**Figure 3.** Validation of the solubility of a) $H_2$, b) $CH_4$, c) $N_2$, and d) $CO_2$ in water at 338.15 K. The solid curves were simulated using the GERG-2008 EoS with calibrated parameters, whereas the scatter markers are experimental data from the literature [12,51–55].

## 2.2. Governing equations and numerical modeling

Generalized mass conservation equations are employed to describe the three-component (i.e., water, hydrogen, and cushion gas), two-phase (i.e., aqueous and gaseous) system. The governing equation is described as follows:

$$\frac{\partial}{\partial t}\left(\emptyset \sum_{\alpha=1}^{n_{ph}} x_{c,\alpha}\rho_\alpha S_\alpha\right) + \nabla \cdot \left(\sum_{\alpha=1}^{n_{ph}} x_{c,\alpha}\rho_\alpha u_\alpha\right) - \sum_{\alpha=1}^{n_{ph}} x_{c,\alpha} q_\alpha = 0 \tag{17}$$

where $\alpha$ represents the phase (aqueous and gaseous), and $c$ denotes the component (hydrogen, water, and cushion gas). $\emptyset$ is the formation porosity. $\rho_\alpha$, $S_\alpha$, and $q_\alpha$ denote the density, saturation, and source flow rate of phase $\alpha$, respectively. $x_{c,\alpha}$ is the mole fraction of component $c$ in phase $\alpha$. $u_\alpha$ is the Darcy velocity of phase $\alpha$, calculated as:

$$u_\alpha = -\frac{kk_{r\alpha}}{\mu_\alpha}\nabla(p_\alpha - \rho_\alpha g) \tag{18}$$

where $k$ is the permeability; $k_{r\alpha}$, $\mu_\alpha$, and $p_\alpha$ are the relative permeability, viscosity, and pressure of phase $\alpha$, respectively; $g$ is the gravitational acceleration. The difference between aqueous phase pressure, $p_{aq}$, and gaseous phase pressure, $p_g$, is referred to as the capillary pressure, $p_c$:

$$p_c = p_{aq} - p_g \tag{19}$$

In this study, the capillary pressure curve is modeled using the Leverett J-function. The relative permeability curves are modeled using the Corey model with the exponent $n$ equal to 2:

$$k_{rnw} = k_{rnw,max} \cdot \left(\frac{S_w}{1-S_{wr}}\right)^2 \tag{20}$$

$$k_{rw} = k_{rw,max} \cdot \left(1 - \frac{S_w}{1-S_{wr}}\right)^2 \tag{21}$$

where $S_{wr}$ is the residual water saturation, equal to 0.15 in this study; $k_{rnw}$ and $k_{rw}$ are the relative permeability of the non-wetting and wetting fluids, respectively; $k_{rnw,max}$ is the maximum non-wetting fluid relative permeability, set as 0.25; $k_{rw,max}$ is the maximum wetting fluid relative permeability, set as 1. **Figure 4** illustrates the capillary pressure and relative permeability curves used in this study, which represent a typical hydrogen-brine-rock system and were measured in the laboratory with a surface contact angle of 34° and interfacial tension of 0.046 N/m [56]. The rock surface wettability, represented by the contact angle, and interfacial tensions have a great influence on the UHS because they determine the capillary pressure, relative permeability, and residual hydrogen saturation [13,57]. No hysteresis in relative permeability is considered in this study because this information was not provided in the experimental measurements.

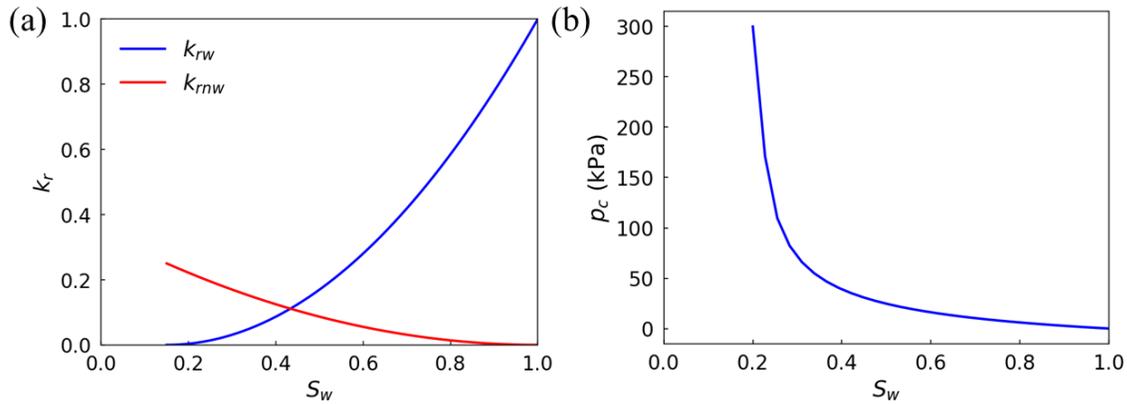

**Figure 4.** a) Relative permeability, and b) capillary pressure curves used in this study.

The stability check is performed to determine the number of phases. If all components are in the aqueous phase, the following relation should hold:

$$\sum_{c=1}^{n_c} z_c K_c - 1 < 0 \tag{22}$$

where $z_c$ is the total mole fraction of component $c$, and $K_c$ is the gaseous-aqueous phase equilibrium ratio calculated as:

$$K_c = \frac{\varphi_{caq}}{\varphi_{cg}} \tag{23}$$

where $\varphi_{caq}$ and $\varphi_{cg}$ are the aqueous and gaseous phase fugacity coefficients for component $c$, respectively.

If there exists only a gaseous phase, then the following relation must hold:

$$\sum_{c=1}^{n_c} \frac{z_c}{K_c} - 1 < 0 \tag{24}$$

which suggests that the system pressure is under the dew point pressure.

When the pressure is between the dew point and bubble point pressures, the system is in the two-phase region; under this condition flash calculations need to be performed. In this work, the Rachford-Rice equation is applied for the pressure-temperature phase equilibrium flash calculation:

$$\sum_{c=1}^{n_c} \frac{z_c(K_c-1)}{1+V(K_c-1)} = 0 \tag{25}$$

where $V$ is the mole fraction of the mixture in the gaseous phase. When the system is in the two-phase region, $x_{c,\alpha}$ is calculated as:

$$x_{c,aq} = \frac{z_c}{1+V(K_c-1)} \tag{26}$$

$$x_{c,g} = K_c x_{c,aq} \tag{27}$$

where $x_{c,aq}$ and $x_{c,g}$ are the mole fractions of component $c$ in the aqueous and gaseous phases respectively.

The overall-composition variables are applied in this study as the primary variables [58]. If the system has $m$ components, then the primary variables will be one phase pressure and $m$-1 total mole fraction, $z_c$. In this study, the aqueous pressure, $p_{aq}$, is used as the primary pressure. The hydrogen's mole fraction ($z_{H_2}$) and one cushion gas's (i.e., $CO_2$, $CH_4$, or $N_2$) mole fraction ($z_{CO_2}, z_{CH_4}$, or $z_{N_2}$) serve as the primary total mole fractions. Based on the proposed primary variables, the expression of the residual functions, $r$, is linearized as:

$$r_c^{v+1} \approx r_c^v + \frac{\partial r_c}{\partial p_{aq}}\bigg|^v \delta p_{aq}^{v+1} + \sum_{i=1}^{m-1} \frac{\partial r_c}{\partial z_i}\bigg|^v \delta z_i^{v+1} = 0 \tag{28}$$

where $v$ and $v + 1$ are two consecutive iteration steps. A consistent fully-implicit discretization scheme is formed using the derivatives of the aqueous phase pressure and total component mole fractions. The derivatives of the phase compositions ($x_{c,\alpha}$) with respect to the total component mole fractions are calculated by the phase equilibrium equations. The primary variables are updated by solving the linear system:

$$J^v \delta \xi^{v+1} = -\zeta^{v+1} \tag{29}$$

where $J$ represents the Jacobian matrix, $\delta \xi^{v+1}$ denotes the Newton's update for the primary variables, and $\zeta^{v+1}$ is the residual equations in the $v + 1$ time step. The linear system with the combination of three components (1, 2, 3) and two phases ($g, aq$) is solved in the form of:

$$\begin{bmatrix} \frac{\partial r_1}{\partial p_l} & \frac{\partial r_1}{\partial z_1} & \frac{\partial r_1}{\partial z_2} \\ \frac{\partial r_2}{\partial p_l} & \frac{\partial r_2}{\partial z_1} & \frac{\partial r_2}{\partial z_2} \\ \frac{\partial r_3}{\partial p_l} & \frac{\partial r_3}{\partial z_1} & \frac{\partial r_3}{\partial z_2} \end{bmatrix} \begin{bmatrix} \delta p_l \\ \delta z_1 \\ \delta z_2 \end{bmatrix} = - \begin{bmatrix} r_1 \\ r_2 \\ r_3 \end{bmatrix} \quad (30)$$

In this study, $r_1$ represents the residuals of the cushion gas (i.e., $CO_2$, $CH_4$, or $N_2$), $r_2$ represents the residuals of hydrogen, and $r_3$ represents the residuals of water. After the system is updated, the stability check is performed using Equation (22) and Equation (24). If the system is in a two-phase region, flash calculations will be performed to determine the mole fractions in each phase using Equations (25) – (27).

## 2.3. Model validation

The proposed multi-component reservoir simulator was validated against Oldenburg's modeling results [59]. The dimensions of the reservoir model are 22 m in depth (Z direction) and 1000 m in length (X direction), as illustrated in **Figure 5A**. We used a mesh grid of 200 × 22 to discretize the reservoir in the x and z directions, respectively. An injection well was positioned at the left boundary of the reservoir, three meters below the top boundary. No-flow boundary condition was applied over all boundaries. The initial pressure of the system was 6 MPa. The reservoir was initially saturated with a pore fluid having 0.8 mole fraction of $CO_2$ and 0.2 mole fraction of water under the temperature of 313.15 K. The simulation was run for 180 days with an $CH_4$ injection rate of $1.8375 \times 10^{-2}$ kg·s$^{-1}$. **Table 3** shows the parameters used in the reservoir simulation.

The midline, defined as the horizontal line located at the middle depth of the reservoir, is used as the reference to measure the mixing zone extent. The mixing middle point, defined as the center between 0.1 mole fraction of $CH_4$ and 0.9 mole fraction of $CH_4$ on the midline, was determined accordingly. The distance from the mixing middle point to the left boundary is defined as the displacement distance, $d_m$. Figure 5A illustrates the midline and the displacement distance. **Figure 5B, C, D** illustrates the simulated spatial distribution of $CH_4$ and $CO_2$ and the development of the mixing zone. The width of the mixing zone increased with time due to the miscible interface between the two gases.

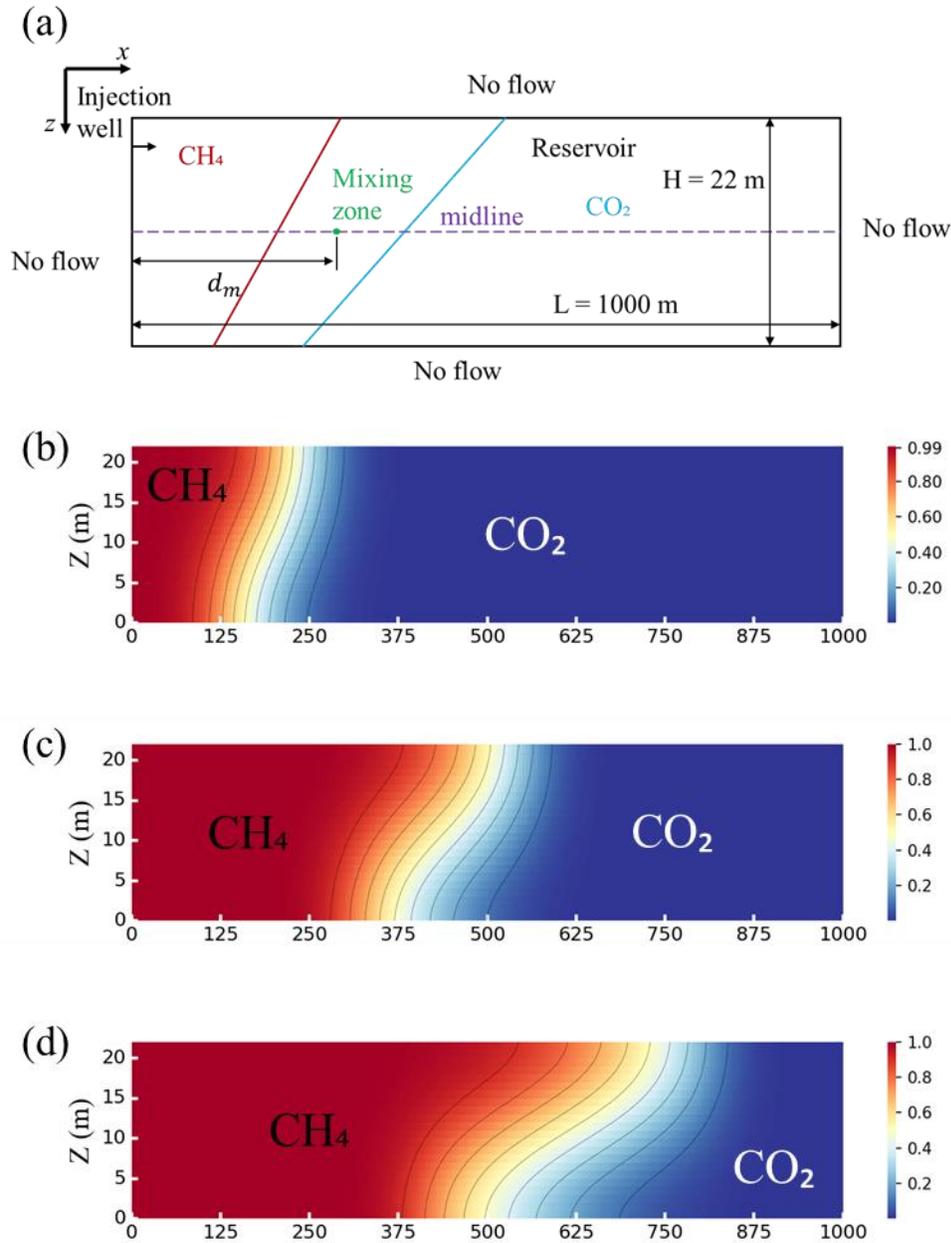

**Figure 5.** a) Schematic diagram of the gas reservoir model. CH$_4$ is the injection gas and CO$_2$ is the cushion gas and simulation results of gas mixing at b) 30, c) 90, and d) 180 days after CH$_4$ injection.

**Table 3.** Properties of the gas storage reservoir.

| Properties | Value | Unit |
| --- | --- | --- |
| Reservoir length | 1000 | $m$ |
| Reservoir height | 22 | $m$ |
| Porosity | 0.3 | - |

| Absolute permeability | $1 \times 10^{-12}$ | $m^2$ |
| temperature | 313.15 | K |
| Initial pressure | 6 | MPa |
| Injection rate | $1.8375 \times 10^{-2}$ | $kg \cdot s^{-1}$ |

**Figure 6A** illustrates the simulated temporal development of the displacement distance. Comparison with the Oldenburg's model showed a good agreement. Both models demonstrated an increase in $d_m$ with time, associated with a decrease in the rate of increase, indicating a decreasing growth rate of the displacement distance. **Figure 6B** illustrates the simulated temporal evolution of the inlet pressure, as well as the comparison with the Oldenburg's model. Both models showed that with the injection of CH$_4$ the inlet pressure increased linearly with time.

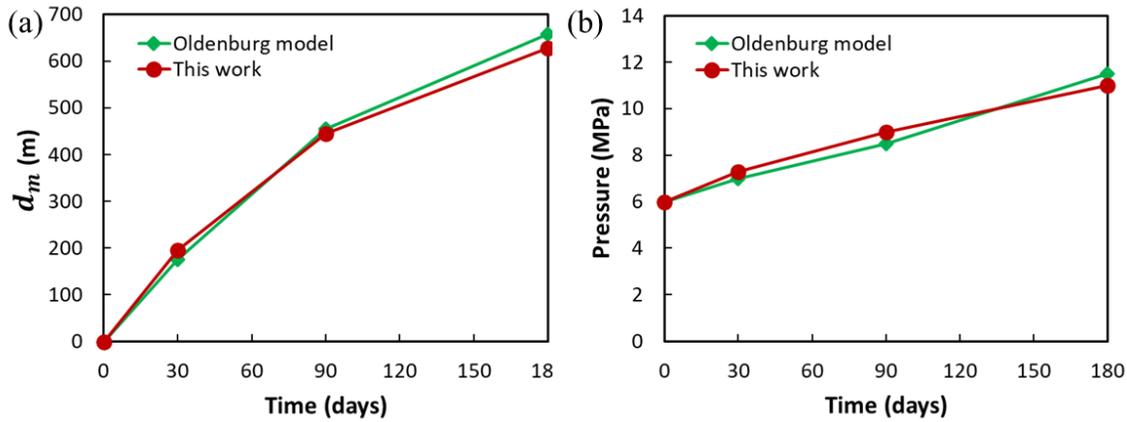

**Figure 6.** a) Displacement distance, and b) pressure at the injection well as a function of time.

## 2.4. Reservoir geometry and operation strategy

**Figure 7A** illustrates the aquifer that was used as the UHS system with various cushion gases. The aquifer has a length of 100 m and height of 50 m, which was discretized into 100 blocks in the x direction and 50 blocks in the z direction with a uniform block size of 1 m. Previous studies have conducted grid size sensitivity analysis in two-phase flow modeling [60,61]. The results showed a marginal difference between different mesh discretization scenarios when the block size was equal to or less than one meter. Hence, it was reasonable to select one meter as the uniform grid size in this study, which reduced the computation time with a negligible numerical error. The injection and production wells were located at the domain center, three meters below the top boundary. The aquifer was initially saturated with water. The parameters and initial conditions are illustrated in **Table 4**. These reservoir parameters were chosen to represent an aquifer at a depth of 2500 m.

**Figure 7B** illustrates the injection and production cycles. The overall simulation time was 20,000 days. During the first 2000 days, a cushion gas was injected at a constant flow rate of $2 \times 10^{-5}$ reservoir pore volume (RPV) per day, equal to 5 normal cubic meters per day (Nm³/day). Next, hydrogen injection and fluid production cycles started, with the hydrogen injection rate and fluid production rate both fixed at $2 \times 10^{-6}$ RPV per day, which was equal to 0.5 Nm³/day. The hydrogen injection and fluid production cycles were conducted alternately through the end of the simulation.

In this study, a dimensionless time, $t^*$, defined as the elapsed time t divided by the characteristic time $t_c$, was used to describe the simulation process:

$$t^* = \frac{t}{t_c} \tag{31}$$

The characteristic time $t_c$ in Equation (31) is calculated as

$$t_c = \frac{L}{u_c} \tag{32}$$

where L is the reservoir length in the x direction and $u_c$ is the characteristic velocity calculated as

$$u_c = \frac{Q}{\varphi H} \tag{33}$$

where $\varphi$ is the porosity of the aquifer, $H$ is the aquifer height, and $Q$ is the cushion gas injection flow rate. Therefore, $t_c$ is equal to $5 \times 10^4$ days in this case. The gaseous and aqueous phase mole fraction profiles are visualized at four different dimensionless times of $t^* = 0.1, 0.2, 0.3$, and 0.4, which corresponded to approximately 5,000, 10,000, 15,000, and 20,000 days, respectively.

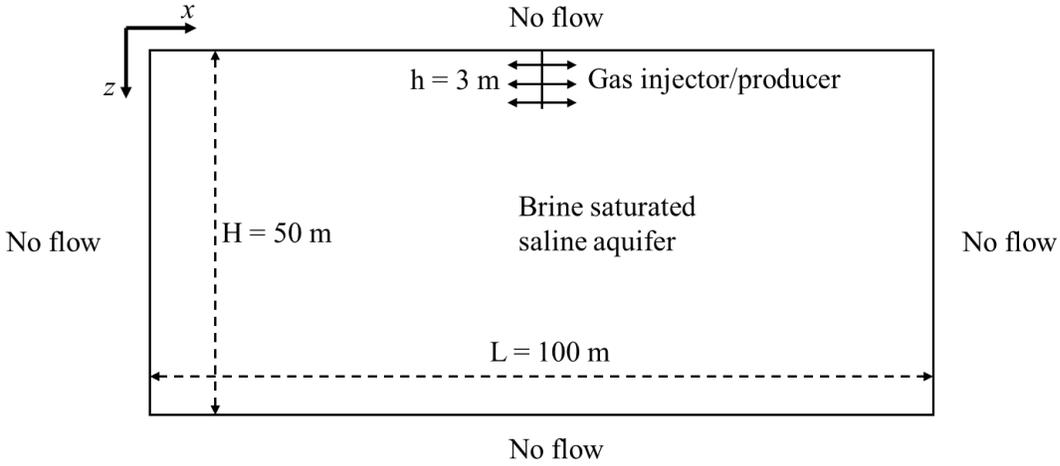

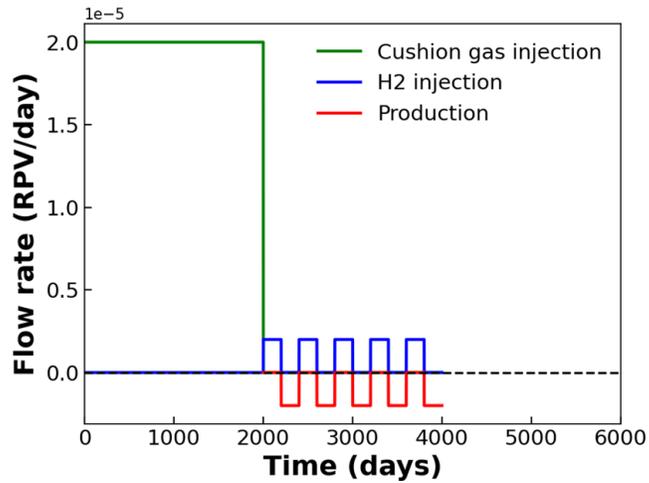

**Figure 7.** a) Schematic diagram of the two-dimensional aquifer and the boundary conditions and b) injection and production cycles.

**Table 4.** Reservoir parameters of the aquifer.

| Parameter | Value | Unit |
|---|---|---|
| Aquifer length | 100 | m |
| Aquifer height | 50 | m |
| Porosity | 0.25 | - |
| Absolute permeability | 120 | mD |
| Initial pressure | $2.5 \times 10^7$ | Pa |
| Bottom hole pressure | $2.5 \times 10^7$ | Pa |
| Temperature | 338.15 | K |

## 3. Results and Discussion

### 3.1. UHS with CO₂ as the cushion gas

**Figure 8** illustrates the spatial distribution profiles of the mole fraction of $CO_2$ and hydrogen in the gaseous and aqueous phases at varying dimensionless times. $CO_2$ spread rapidly in the lateral direction in both the gaseous and aqueous phases, as shown in Figures 8A and 8B. The solubility of $CO_2$ in brine led to an increase in the fluid density near the top boundary, which caused gravitational instability and consequently resulted in the downward convection of $CO_2$-rich solution [62–64], as shown in Figure 8B.

With the increase of hydrogen injection and production cycles, the mole fraction of hydrogen in the gaseous phase increased continuously, pushing the $CO_2$ component in the gaseous phase to the left and right sides of the reservoir and consequently creating a transition zone between the two gas components, as demonstrated in Figure 8C. In addition, downward convection of $CO_2$-rich solution enhanced the dissolution of hydrogen in the aqueous phase, because the downward convection led to recirculation flows that carried underlying brine to the aqueous-gaseous phase interface, which was favorable for hydrogen dissolution in the aqueous phase.

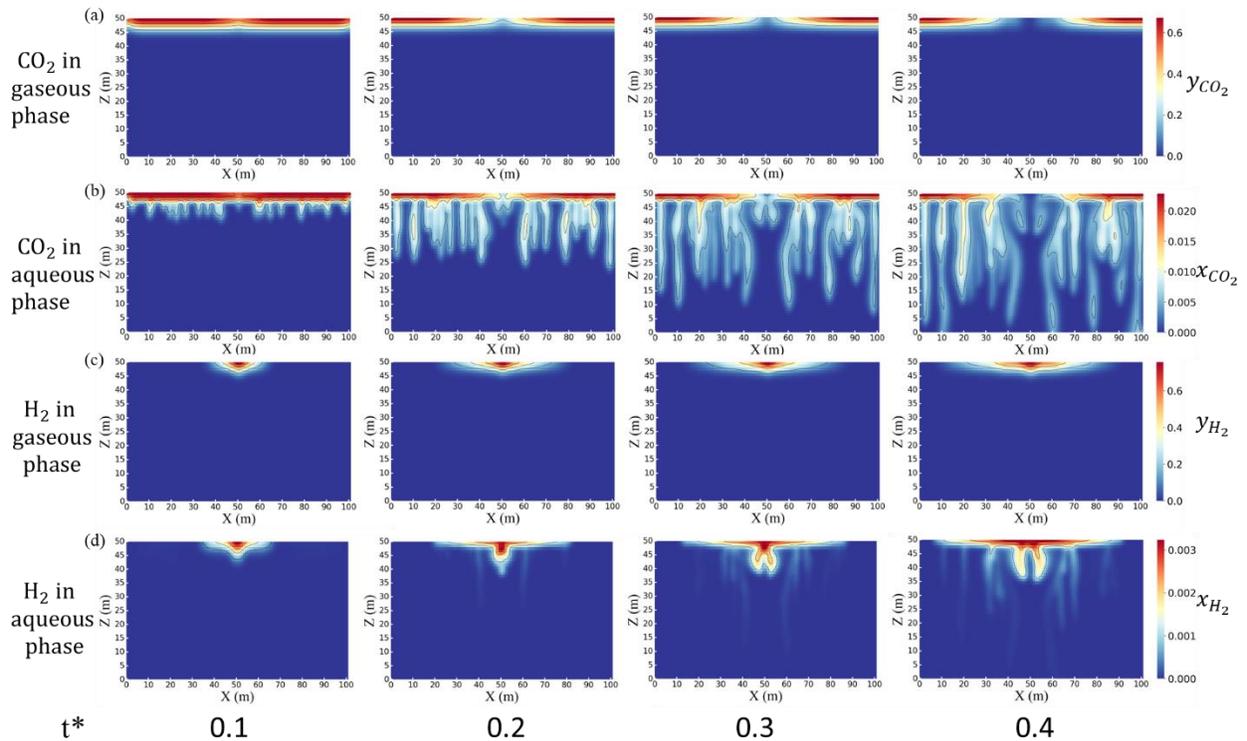

**Figure 8.** Temporal evolution of a) $CO_2$ mole fraction in gaseous phase, b) $CO_2$ mole fraction in aqueous phase, c) $H_2$ mole fraction in gaseous phase, and d) $H_2$ mole fraction in aqueous phase in an aquifer UHS system with $CO_2$ as the cushion gas.

## 3.2 UHS with CH₄ and N₂ as cushion gases

**Figures 9** and **10** illustrate the mole fraction distributions with $CH_4$ and $N_2$ as the cushion gas, respectively. The solubility of $CH_4$ is slightly higher than that of $N_2$, but both are significantly lower than that of $CO_2$. As a consequence, there was no noticeable density-driven downward convection observed in these two scenarios, as shown in Figures 9 and 10. Therefore, $CH_4$ and $N_2$ are ideal cushion gases for UHS because of their low solubility in water, which prevents the loss of $CH_4$ and $N_2$ through density-driven downward convection and consequently keeps injected hydrogen close to the borehole.

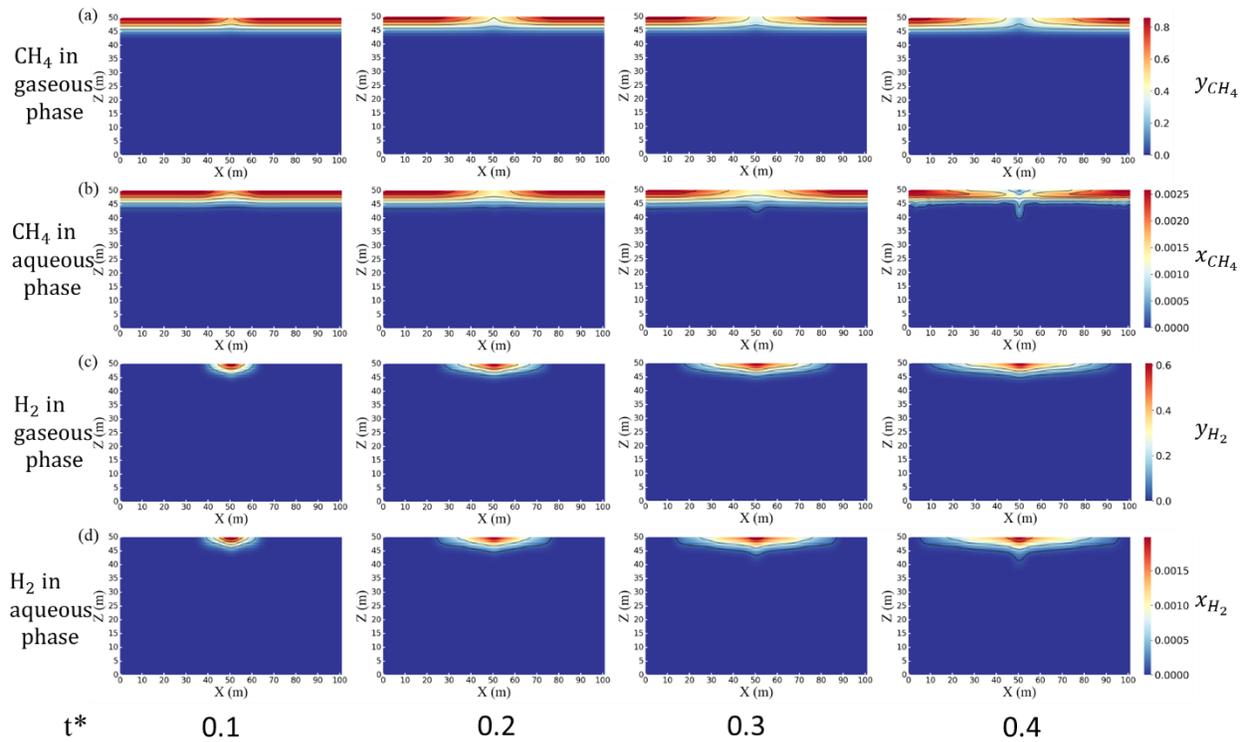

**Figure 9.** Temporal evolution of a) $CH_4$ mole fraction in gaseous phase, b) $CH_4$ mole fraction in aqueous phase, c) $H_2$ mole fraction in gaseous phase, and d) $H_2$ mole fraction in aqueous phase in an aquifer UHS with $CH_4$ as the cushion gas.

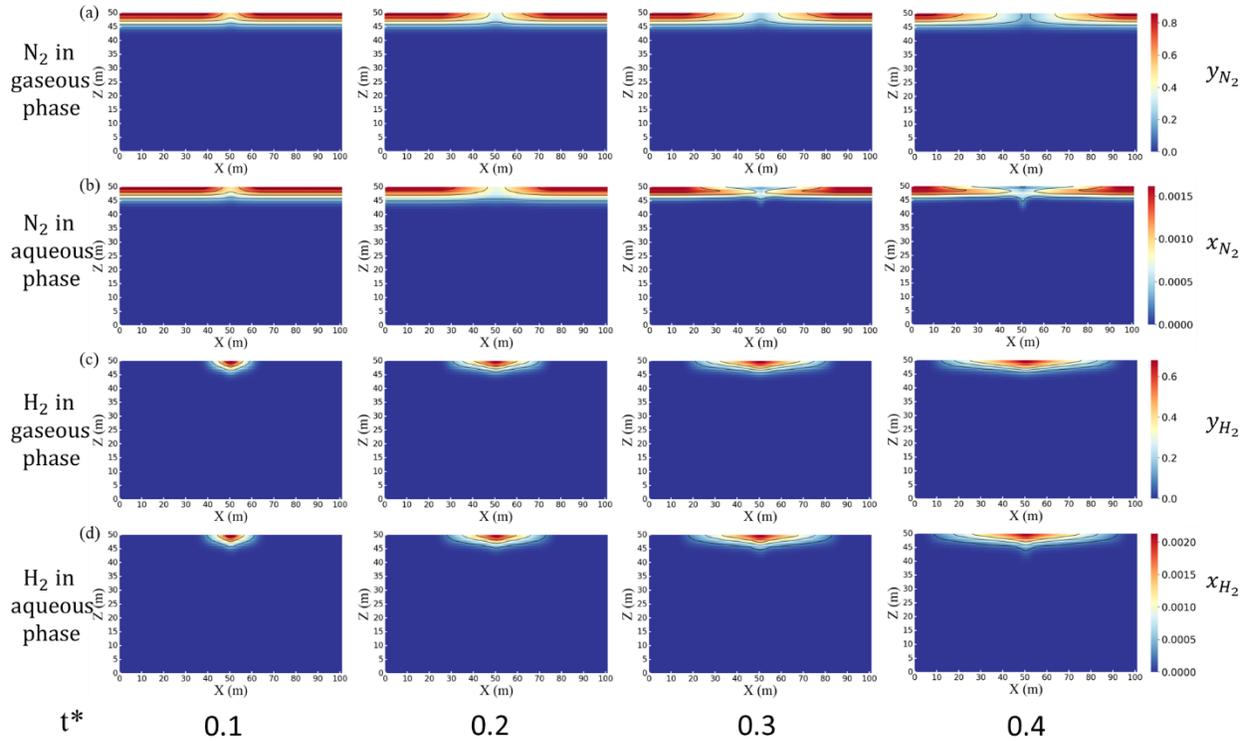

**Figure 10.** Temporal evolution of a) $N_2$ mole fraction in gaseous phase, b) $N_2$ mole fraction in aqueous phase, c) $H_2$ mole fraction in gaseous phase, and d) $H_2$ mole fraction in aqueous phase in an aquifer UHS with $N_2$ as the cushion gas.

### 3.3 Quantitative analysis of the impact of cushion gases

In this section, the performance of UHS with different cushion gases is analyzed quantitatively. **Figure 11** illustrates the gaseous phase saturations and aqueous phase pressures at the borehole. At the beginning of the cushion gas injection period (i.e., the first 2000 days), a substantial increase of the total gaseous saturation can be observed for all the three cases. The injection of the cushion gas displaced the initial pore fluids and expanded the gaseous volume in the upper layer of the aquifer, resulting in a higher relative permeability and mobility of the gaseous phase. A cyclic pattern of increase and decrease in the gaseous saturation can be observed after the first 2000 days, corresponding to the injection and production cycles. Specifically, the total gaseous saturation increased during the hydrogen injection stages and decreased during the production stages. Meanwhile, the overall gaseous saturation in the $CH_4$ and $N_2$ scenarios continuously increased over time but continuously decreased in the $CO_2$ scenario. This difference was attributed to the different solubilities of the cushion gases. $CH_4$ and $N_2$ had low solubility and only a small portion of the cushion gas dissolved in the brine. In the cyclic hydrogen injection and production stage, aqueous phase fluids were extracted, leading to continuously increasing gaseous phase saturation near the wellbore. In contrast, $CO_2$ had a high solubility, leading to a reduction in the gaseous volume and consequently a decrease in the gaseous saturation near the wellbore.

Figure 11B illustrates the aqueous phase pressure at the borehole. The aqueous phase pressure in the $CH_4$ and $N_2$ scenarios were similar. The presence of dissolved $CO_2$ in the aqueous phase enhanced the brine density, resulting in a higher aqueous phase pressure compared to the other two scenarios. The aqueous phase pressure continuously decreased in the cyclic hydrogen injection and production stage in all the three scenarios. This was because aqueous phase fluids were extracted from the aquifer during the production stages, leading to a reduction of the water mass near the borehole.

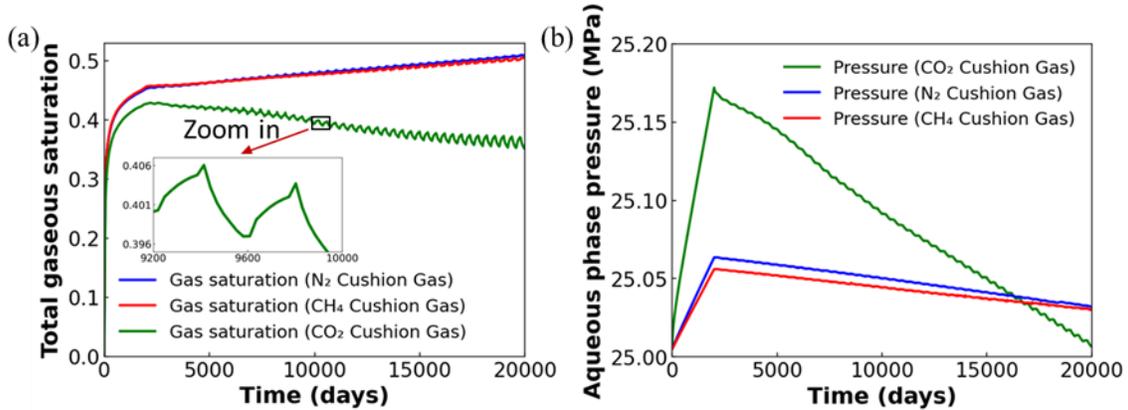

**Figure 11.** Temporal evolution of a) total gaseous phase saturation, and b) aqueous phase pressure at the borehole.

The production rates of the cushion gas and hydrogen in one single production cycle have been investigated in previous studies [25,26,33]. In this study, we focused on the long-term UHS performance associated with multiple injection and production cycles. **Figure 12A, B** illustrates the production rates of hydrogen and cushion gases as a function of time in the cyclic hydrogen injection and production period. The production fluids include both gaseous and aqueous phase components. In this graph, each marker is the production rate averaged over that particular 200 days of production stage. All the volume calculations of the fluids have been converted to the standard temperature and pressure (STP) conditions. The hydrogen production rates in the $CH_4$ and $N_2$ scenarios showed similar patterns. In both scenarios, the hydrogen production rate increased with time whereas the cushion gas production rate decreased with time. This was because the cyclic injection of hydrogen into the aquifer led to an increase in its concentration and a drop in the cushion gas concentration near the borehole.

Conversely, in the scenario with $CO_2$ as the cushion gas, the hydrogen production rate decreased and the $CO_2$ production rate reached zero after 10,000 days. Figure 8A illustrates that the $CO_2$ mole fraction in the gaseous phase was approximately 0.2 at 10,000 days (i.e., $t^* = 0.2$) and approached zero at 15,000 days (i.e., $t^* = 0.3$), indicating that there was no gaseous-phase $CO_2$ produced after 10,000 days. In this scenario, with the decrease in the total gaseous saturation as shown in Figure 11A, the amount of produced water increased. Therefore, the produced hydrogen volume decreased because most hydrogen was in the gaseous phase. This finding shows that the dissolution of $CO_2$ in water may impede the recovery of hydrogen in an UHS system when $CO_2$ serves as the only cushion gas.

**Figure 12C** illustrates the cumulative dissolved hydrogen volumes in the cyclic hydrogen injection and production period as a function of time in the three cushion gas scenarios. The gas volumes were measured under the STP conditions. In the $CH_4$ and $N_2$ scenarios, the dissolved hydrogen volume did not grow noticeably. That was because the dissolution of $CH_4$ and $N_2$ had reached the equilibrium state by the end of the cushion gas injection stage and the dissolved cushion gas barely increased the solution density. As a consequence, there was no density-driven downward convection which could significantly enhance the interface area between the gaseous and aqueous phases. Therefore, hydrogen kept staying at the top of the aquifer. Once the top of the aquifer saturated with hydrogen, there will be no more hydrogen dissolved into the aquifer, leading to limited hydrogen dissolution in these two scenarios.

The solubility behavior of hydrogen in the $CO_2$ scenario was similar to that in the $CH_4$ and $N_2$ scenarios during the first 4000 days. However, more hydrogen dissolved into brine after that and the volume of the dissolved hydrogen increased noticeably with time through the end of the simulation (20,000 days). This was because after the first 5,000 days the $CO_2$-rich brine near the top boundary triggered gravitational instability and consequently density-driven downward convection, which enhanced the interface area between gaseous and aqueous phases and thus was favorable for hydrogen dissolution into the aqueous phase.

**Figure 12D** illustrates the purity of produced hydrogen as a function of time during the cyclic hydrogen injection and production stage. The hydrogen purity is calculated as the ratio of the volume of gaseous phase hydrogen to the volume of the total produced gaseous phase at the surface, which consists of hydrogen and the cushion gas. The produced hydrogen purity in general increased over time in all the three cushion gas scenarios. Particularly, in the $CO_2$ case, the hydrogen purity continued increasing until it reached one. This was because $CO_2$ entirely dissolved into the aqueous phase during the cyclic hydrogen injection and production stage, leading to absence of $CO_2$ near the wellbore. Therefore, at a later time in the cyclic hydrogen injection and production stage, the gaseous phase near the wellbore was entirely composed of hydrogen, leading to a 100% purity of produced hydrogen.

Figure 12D also shows that the purity of produced hydrogen with $N_2$ as the cushion gas was higher than that with $CH_4$ as the cushion gas. This was attributed to the higher density of $N_2$, which resulted in stratification of gas component distribution in the gaseous phase region. Specifically, $N_2$ stayed at the bottom of the gaseous phase due to its higher density, leading to higher hydrogen concentration at the top region and consequently a higher purity of produced hydrogen.

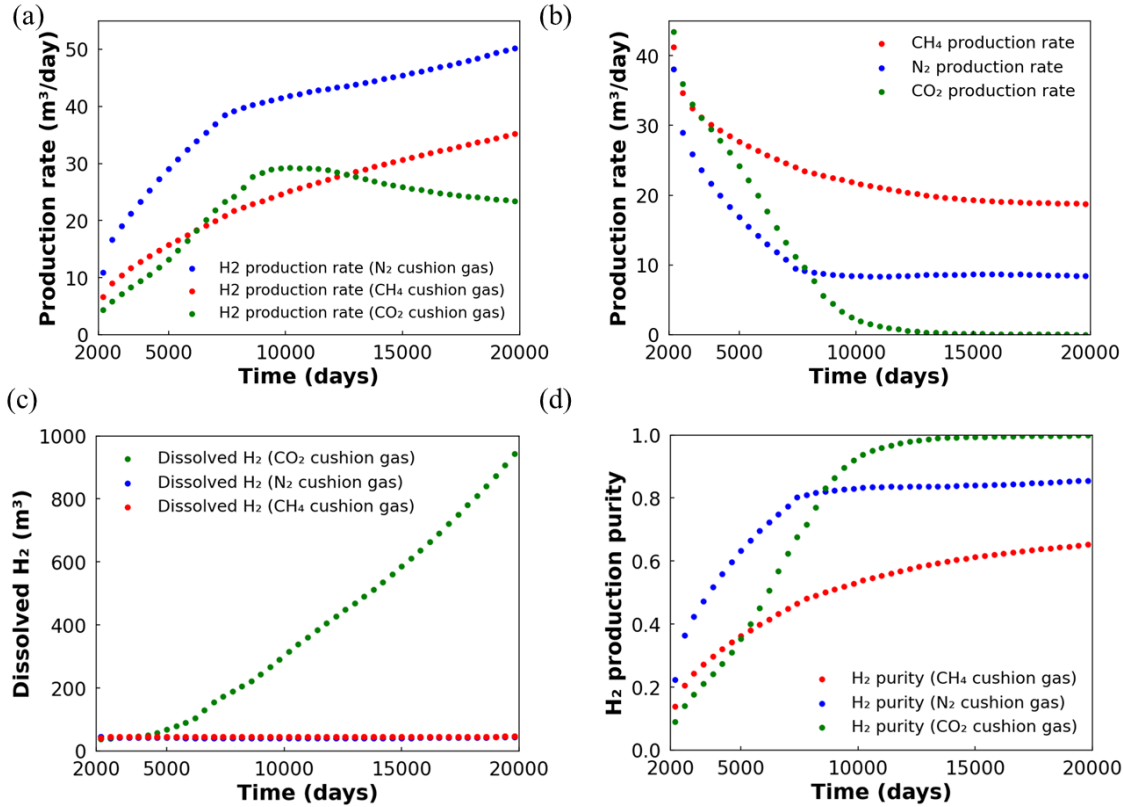

**Figure 12.** Production rate of a) $H_2$, and b) cushion gases in the cyclic $H_2$ injection and production period. c) Dissolved $H_2$ volume with different cushion gases as a function of time. d) Production purity of $H_2$ as a function of time in the cyclic $H_2$ injection and production period.

Based on these simulation results, a cushion gas' thermodynamic properties play a significant role in the hydrogen production purity. Specifically, the density difference between the cushion gas and hydrogen can induce gas stratification, which in turn increases the hydrogen production purity. Additionally, a higher cushion gas solubility increases hydrogen production purity because less cushion gas is present in the gaseous phase. However, this scenario is not necessarily favorable for the UHS due to the associated loss of the cushion gas, which could potentially lead to a decrease in the reservoir pressure.

### 3.4 Comparison with other studies in the literature

There are several studies in the literature investigating the UHS using reservoir simulations. An important factor in reservoir simulation is the operation strategy, which may cause different simulation results. In Okoroafor et al.'s study [34], they used methane as a cushion gas and observed a decrease in hydrogen's production rate and an increase in methane's production rate with increasing injection and production cycles. In contrast, our simulation demonstrated an increasing hydrogen production rate and a decreasing methane production rate. This difference was caused by the different reservoir operation strategies adopted in the two studies. In Okoroafor et al.'s simulation, the production rate was much higher than the injection rate, leaving less and less residual hydrogen in the aquifer with increasing injection and production cycles, thereby causing a decline in the hydrogen production purity. Consequently, our simulation used a production rate equal to the injection rate, resulting in a greater amount of hydrogen remaining in the aquifer. Feldmann et al.'s study [30] is consistent with our simulation results, which showed escalating hydrogen production purity with increasing cycles, likely due to a similar operational strategy involving equal injection and production rates.

The selection of the cushion gas is another significant factor in UHS, with various studies yielding different results regarding the cushion gas performance [23,33]. Zamehrian and Sedaee [23] concluded that the nitrogen injection led to the highest hydrogen recovery, which resulted from the higher pressure in the condensate; this finding is consistent with our simulation results. . Kanaani et la. [33] reported superior performance associated with methane injections, despite the higher average reservoir pressure associated with nitrogen injections. They argued that the less density difference between hydrogen and methane mitigated the gas stratification. Different from Kanaani et al's well configuration [33], the production well location in our model is at the top of the storage aquifer, thereby minimizing hydrogen trapping caused by the gas stratification. Also, the similar densities of methane and hydrogen resulted in a greater extent of gas mixing, leading to a decrease in the hydrogen production purity in our simulation.

Additionally, Kanaani et al.'s [33] work concluded that almost the same amount of hydrogen ended up being trapped in the water phase in the methane, nitrogen, and carbon dioxide scenarios, which is different from our findings. Our thermal dynamic model indicates that the solubility of $CO_2$ is over ten times larger than other cushion gases, which results in significant amounts of $CO_2$ dissolved in the pore water, thereby significantly impacting hydrogen flow and trapping. These findings highlight the complexity of assessing the UHS performance and underscore the pressing needs in systematically integrating cushion gas thermodynamic properties, operational strategy, and reservoir design in future investigations.

## 4. Conclusions and Implications

This study investigated the effects of different cushion gases (i.e., $CO_2$, $CH_4$, and $N_2$) on the performance of an UHS system in a subsurface aquifer. To this end, a two-phase three-component reservoir simulator was developed. The fluid thermodynamic properties, including fluid density, viscosity, and solubility, were calculated using the GERG-2008 EoS, in which the parameters were determined by fitting the model to experimental data. The distribution of the mole fractions of hydrogen and the cushion gas in the gaseous and aqueous phases was simulated over a period of 20,000 days, which included the cushion gas injection stage and the cyclic hydrogen injection and production stage. This study then analyzed and compared the hydrogen production rate, dissolution, and production purity between three cushion gas scenarios.

Simulation results showed that the gaseous phase hydrogen stayed in the top region of the aquifer with $CH_4$ and $N_2$ as the cushion gas. The low solubility of $CH_4$ and $N_2$ led to a higher fraction of gaseous phase staying at the top of the aquifer, which was favorable for maintaining the hydrogen production rate due to the enhanced reservoir pressure but unfavorable for the purity of produced hydrogen. In contrast, hydrogen-rich fingers occurred in the aqueous phase when $CO_2$ was used as the cushion gas, because dissolved $CO_2$ increased the brine density, leading to density-driven downward convection, which enhanced hydrogen dissolution in the aqueous phase. Although the $CO_2$ scenario yielded the highest production purity of hydrogen, the high solubility of $CO_2$ in brine resulted in a decreased gaseous phase saturation, thereby reducing the gaseous hydrogen mobility near the wellbore. Therefore, the selection of the cushion gas should be made after careful considerations of the specific geological conditions and desired project outcomes.

This work is the first study that utilizes an EoS-based reservoir simulator to investigate hydrogen's flow patterns and interactions with cushion gases in an underground storage system. Particularly, the effects of different cushion gases (i.e., $CO_2$, $CH_4$, and $N_2$) on the performance of the UHS were quantitatively investigated. The developed reservoir simulation tools and research findings from this study will be valuable to support decision making in practical UHS projects.

**Declaration of Competing Interest**

The authors declare that they have no known competing financial interests or personal relationships that could have appeared to influence the work reported in this paper.

**Data availability**

Data will be made available on request.

**Acknowledgement**

The authors would like to thank Stevens Institute of Technology for the financial support of a startup package. The authors are thankful to the Delft Advanced Reservoir Simulation (DARSim) research group for their valuable assistance and support. We sincerely appreciate the insightful suggestions and guidance offered by Dr. Hadi Hajibeygi, whose constructive feedback significantly enhanced the quality of this work. Dr. Yuhang Wang was supported by the "CUG Scholar" Scientific Research Funds (Project No. 2022157) at the China University of Geosciences (Wuhan).